\documentclass[aps,prl,superscriptaddress,showpacs,twocolumn,floatfix]{revtex4}
\usepackage{graphicx}
\usepackage[usenames,dvipsnames]{color}

\newcommand{\gevsq}{GeV$^2$}

\newcommand{\qsq}{$Q^2$}

\newcommand{\GEp}{$ {G_{_{\rm E}}^{\it p}}$}
\newcommand{\GMp}{$ {G_{_{\rm M}}^{\it p}}$}
\newcommand{\GEn}{$ {G_{_{\rm E}}^{\it {\,n}}}$}
\newcommand{\GMn}{$ {G_{_{\rm M}}^{\it \,n}}$}

\newcommand{\Fone}{${ F_1}$}
\newcommand{\Ftwo}{${ F_2}$}
\newcommand{\Fonep}{\mbox{${ {F_1^{\it p}}}$}}
\newcommand{\Ftwop}{\mbox{${ {F_2^{\it p}}}$}}
\newcommand{\Fonen}{\mbox{${ {F_1^{\it n}}}$}}
\newcommand{\Ftwon}{\mbox{${ {F_2^{\it n}}}$}}
\newcommand{\Foneu}{\mbox{${ {F_1^{\it u}}}$}}
\newcommand{\Ftwou}{\mbox{${ {F_2^{\it u}}}$}}
\newcommand{\Foned}{\mbox{${ {F_1^{\it d}}}$}}
\newcommand{\Ftwod}{\mbox{${ {F_2^{\it d}}}$}}
\newcommand{\Sp}{${ {Q^2 F_2^{\it p}/F_1^{\it p}}}$}
\newcommand{\Sn}{${ {Q^2 F_2^{\it n}/F_1^{\it n}}}$}

\newcommand{\PR}{{ Phys. Rev. }}
\newcommand{\PRL}{{ Phys. Rev. Lett. }}

\newcommand{\etal}{{\em et al.}}
\newcommand\UVa{University of Virginia, Charlottesville, VA 22903}
\newcommand\UMASS{University of Massachusetts, Amherst, MA 01003}
\newcommand\JLab{Thomas Jefferson National Accelerator Facility,
  Newport News, VA 23606 }
\begin{document}

\preprint{APS/123-QED}

\title{Flavor decomposition of the elastic nucleon electromagnetic form factors}
\author{G.D.~Cates}          \affiliation{\UVa}
\author{C.W.~de~Jager}     \affiliation{\JLab}
\author{S.~Riordan}           \affiliation{\UMASS}
\author{B.~Wojtsekhowski} \thanks{Corresponding author: bogdanw@jlab.org} 
\affiliation{\JLab}

\date{\today}

\begin{abstract}
The $u$- and $d$-quark contributions to the elastic nucleon electromagnetic
form factors have been determined using experimental data on \GEn, \GMn, \GEp, and \GMp.
Such a flavor separation of the form factors became possible
up to 3.4~\gevsq\ with recent data on \GEn\ from Hall A at JLab.
At a negative four-momentum transfer squared $Q^2$ above 1~\gevsq, 
for both the $u$- and $d$-quark components, the ratio of the Pauli form factor
to the Dirac form factor,  \Ftwo/\Fone, was found to be almost constant, 
and for each of \Ftwo\ and \Fone\ individually, the $d$-quark portions
of both form factors drop continuously with increasing \qsq.

\end{abstract}
\pacs{14.20.Dh, 13.40.Gp, 24.70.+s, 25.30.Bf}
\maketitle

Electron-nucleon scattering has been extensively studied in two cases.
The first case is in elastic scattering which is characterized by the electromagnetic 
form factors~\cite{ros50}.
The second case is in deep inelastic scattering (DIS) characterized by
the structure functions which exhibit Bjorken scaling~\cite{bjo69}.
The study of the proton form factors in elastic scattering by
Hofstadter {\it et al.} provided some of the first information on
the size of the proton and the distribution of charge and magnetization~\cite{hof56}.  
Deep inelastic scattering resulted in the discovery of quarks at the
Stanford Linear Accelerator Center (SLAC) ~\cite{bre69}, and also taught us 
about the nucleon's spin structure~\cite{spins}. 
Elastic and deep inelastic scattering provide complementary information: elastic scattering
reveals features of transverse structure,  and DIS provides information
regarding the longitudinal momentum distributions of the quarks.
Taken together, elastic and the deep inelastic scattering can both be understood within the 
broader framework of generalized parton distributions which facilitate a 
tomographic picture of the nucleon~\cite{gpds1}.

Experimental data on the proton Dirac form factor $F_1^p$ ~\cite{arn86} 
have been found to be in fair agreement with a scaling prediction 
based on perturbative QCD (pQCD), \Fonep$\, \propto Q^{-4}$~\cite{lep79}, 
where $Q^2$ is the negative four-momentum transfer squared.
However, it has been argued that pQCD is not applicable for exclusive processes 
at experimentally accessible values of momentum transfer~\cite{isgur}.
Indeed, experimental results from Thomas Jefferson Laboratory 
(JLab)~\cite{jon00} for the ratio of the proton Pauli form
factor \Ftwop\ and the Dirac form factor \Fonep\ have been found to be 
in disagreement with the suggested scaling $F_2^p/F_1^p \propto 1/Q^2$~\cite{lep79}.  
These same data, however, are in reasonable agreement with an updated 
pQCD prediction \qsq\Ftwo/\Fone$\propto \ln^2 [Q^2/\Lambda^2]$~\cite{bel03}
even at modest \qsq\ of several \gevsq.
Here $\Lambda$ is a soft scale parameter related to the size of the nucleon. 
The prediction has the important feature that it includes components of the
quark wave function with nonzero orbital angular momentum.

In view of these facts it is of significant interest to look for the
origin of the observed \qsq-dependence of \Ftwop/\Fonep.
We report here on the flavor-separated form factors for the up and down quarks up 
to \qsq$\,=\,$3.4~\gevsq.  
When considering the ratio \Ftwo/\Fone\ for the $d$ and $u$-quark 
contributions to the nucleon form factors, we find their
\qsq\ dependencies to be surprisingly constant.
However, when combined in the proton
form factors, they give the appearance of the onset of scaling.
It is interesting to note that the authors of~\cite{bel03} did not expect the asymptotic 
predictions for the form factors to work at a few \gevsq\ and considered that 
`` ... the observed consistency might be a sign of precocious scaling as 
a consequence of delicate cancellations in the ratio''.

In the one-photon exchange approximation, the amplitude for
electron-nucleon elastic scattering can
be written $ {M^{\rm^{EM}} = -(4\pi\alpha/Q^2)l^{\mu}\,J_{\mu}^{\rm^{EM}}}$,
where $\alpha$ is the fine structure constant,  
$l^{\mu}=\overline{e} \gamma^{\mu} e$ is the leptonic vector current, and 
\begin{equation}
J_{\mu}^{\rm^{EM}} = \langle p (n) |({\textstyle{2\over3}}\overline{u}\gamma_{\mu}u 
+ {\textstyle{{-1\,}\over{\ 3}}}\overline{d}\gamma_{\mu}d)| p (n) \rangle
\label{eq:jmu}
\end{equation}
is the hadronic matrix element of the electromagnetic current operators for the proton (neutron).
Here, the neglect of heavier quarks in this context is supported by experimental data 
on parity non-conserving polarized electron scattering from the
proton~\cite{pvgr}.
While we can not evaluate the matrix elements of
 $\overline{u}\gamma_{\mu}u$ and $\overline{d}\gamma_{\mu}d$ explicitly,  
from symmetry considerations we know that the matrix element shown in
Eq.~\ref{eq:jmu} must have the form (considering the proton for definiteness)
\begin{equation}
J_{\mu}^{\rm^{EM}} = \overline{p}(k^{\prime})
\left[ \gamma^{\mu} F^p_1(Q^2) + {{i\sigma^{\mu\nu}q_{\nu}}\over{2M}} F^p_2(Q^2) \right] p(k),
\end{equation}
where $p(k)$ and $\overline{p}(k^{\prime})$ are the proton Dirac spinors 
for the initial and final momenta $k$ and $k^{\prime}$, respectively.
The definition of the neutron form factors \Fonen(\qsq) and \Ftwon(\qsq) follows similarly.

\begin{figure}[!hbt]
\includegraphics[trim = 5mm 5mm 5mm 5mm, width=0.45\textwidth]{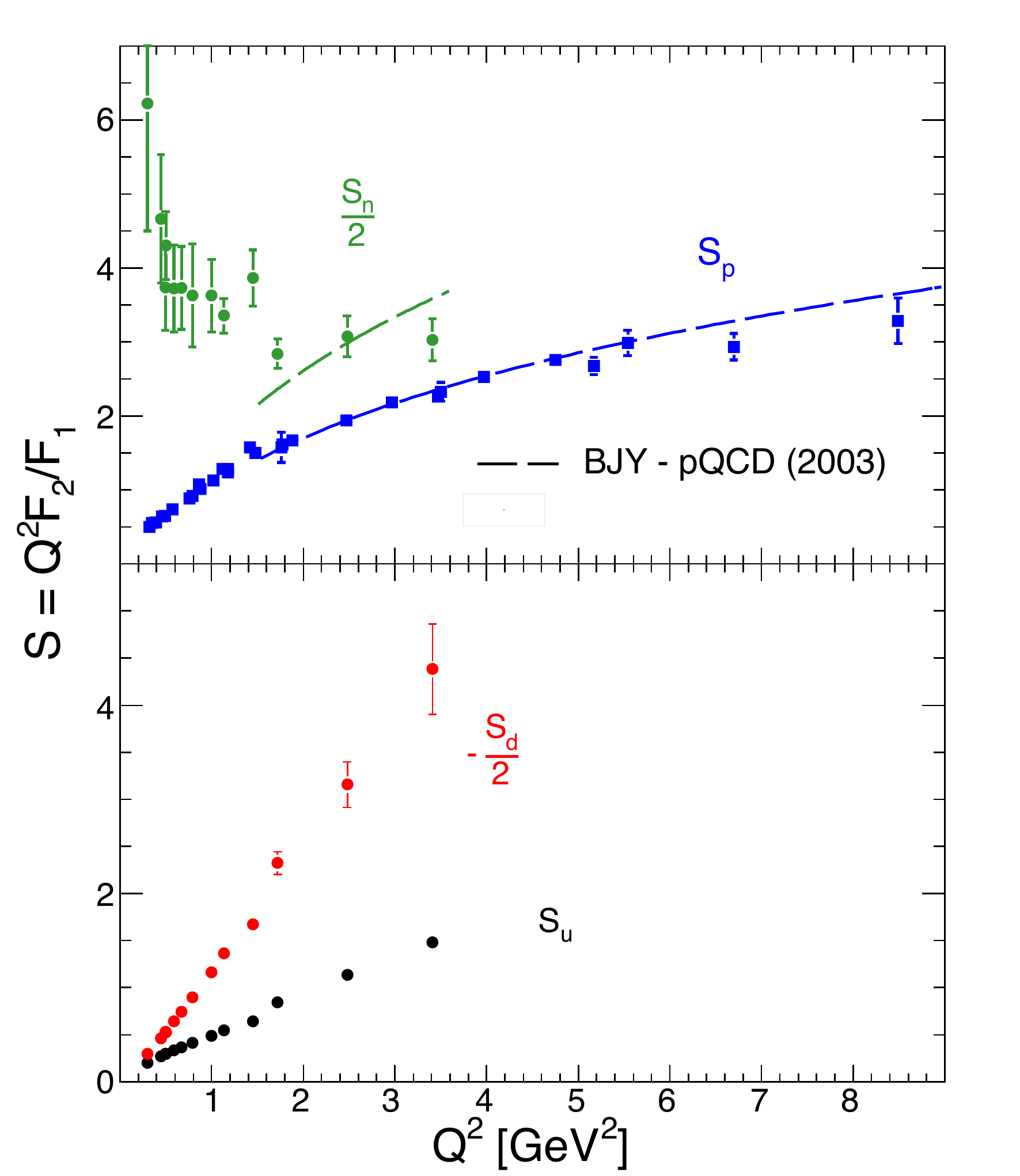}
\caption
{The ratio of the Pauli and Dirac form factors, multiplied by \qsq, $S=$\qsq\Fone/\Ftwo, vs.
the negative four-momentum transfer squared \qsq.
The upper panel shows $S_p$ for the proton and $S_n$ for the neutron using data 
from Refs.[13-18], as well as the curves of the prediction~\cite{bel03}: 
$\ln^2[Q^2/\Lambda^2]$ for $\Lambda$=300 MeV which is normalized to the data at 2.5~\gevsq.
The bottom panel shows the individual flavor quantities $S_u$ and $S_d$ for 
the $u$ and $d$ quarks, respectively.}
\label{fig:q2f2f1} 
\vskip -.15in
\end{figure} 
The JLab data for \GEp/\GMp\ from Refs.~\cite{jon00} were used to plot
$S_p \, \equiv \,$\Sp\ in the upper panel of Fig.~\ref{fig:q2f2f1},
which also shows the prediction~\cite{bel03} at $\Lambda$ = 300 MeV. 
Data on \GEn/\GMn\ for the neutron up to  \qsq=3.4~\gevsq{}
were recently published by Riordan \etal~\cite{rio10}. 
For the first time, it is possible to examine the behavior of the neutron ratio \Ftwon/\Fonen\
in the same \qsq{} range as that where the interesting behavior was first seen for the proton~\cite{jon00}. 
Using the data of Riordan~\etal{} as well as those of Refs.[14-18], we
also show in Fig.~\ref{fig:q2f2f1} the quantity $S_n \, \equiv \,$\Sn.  
Scaling of $S_n$ is clearly not evident at the lower \qsq\
values shown, although the data do not rule out this type of behavior
at a moderately higher \qsq.

Thus far, by discussing $F_1^{\it p(n)}$ and $F_2^{\it p(n)}$ we are explicitly examining
the behavior of the matrix element of the electromagnetic operators
$({\textstyle{2\over3}}\overline{u}\gamma_{\mu}u + {\textstyle{{-1\,}\over{\ 3}}}\overline{d}\gamma_{\mu}d)$
in the proton (neutron).  
If we assume charge symmetry
(thus implying $\langle p |\overline{u}\gamma_{\mu}u|p\rangle = \langle n
|\overline{d}\gamma_{\mu}d| n \rangle$),
it is possible to perform a flavor decomposition of the form factors $F_1^{\it p(n)}$ and $F_2^{\it p(n)}$,
and construct form factors corresponding to the matrix elements of $\overline{u}\gamma_{\mu}u$ and 
$\overline{d}\gamma_{\mu}d$ individually~\cite{mil90}.  
Here we use the relations
\begin{center}
$F_{1(2)}^u \, = \, 2\,F_{1(2)}^p \,+\, F_{1(2)}^n$
  \hskip .05in and \hskip .05in 
$F_{1(2)}^d \, = \, 2\,F_{1(2)}^n \,+\, F_{1(2)}^p$. \\
\end{center} 
In what follows, we use the convention that $F_{1(2)}^u$ and
$F_{1(2)}^d$ refer to the up and down quark contributions 
to the Dirac (Pauli) form factors of the proton.
At \qsq=0, the normalizations of the Dirac form factors are given by: $F_{1}^{u}(0) = 2$ ($F_{1}^{d}(0) = 1$) 
so as to yield the normalization of 2 (1) for the $u$ ($d$)-quark distributions in the proton. 
The normalizations of the Pauli form factors at \qsq=0 are given by $F_{2}^{q}(0) = \kappa_q$, where 
$\kappa_u$ and $\kappa_d$ can be expressed in terms of the proton ($\kappa_p$) 
and neutron ($\kappa_n$) anomalous magnetic moments as
\begin{center}
$\kappa_u \,\equiv \, 2 \kappa_p + \kappa_n =+1.67$ \hskip .05in and \hskip .05in
$\kappa_d \,\equiv \, \kappa_p + 2 \kappa_n =-2.03$.
\end{center}

Having defined the flavor-separated Dirac and Pauli form factors, we can also define the quantities
\begin{center}
$S_u \equiv Q^2F_2^{\it u}/F_1^{\it u} $ \hskip .25in and \hskip .25in
$S_d \equiv Q^2F_2^{\it d}/F_1^{\it d} $, \\
\end{center} 
which we have plotted in the bottom panel of Fig.~\ref{fig:q2f2f1}.  
Each individual data point corresponds to an experimental
result on $G_E^n/G_M^n$ from Refs.[13-18].
Only the uncertainties in the ratio \GEn/\GMn\ are included in the
error bars of the flavor-separated results because the other form factors 
(calculated with the Kelly fit~\cite{kel04}) are known to much higher
accuracy, albeit dependent on the particular parameterization chosen.
The behavior we see is completely different from that of the proton
and the neutron.  
There is a striking lack of saturation, and indeed the variation of
$S_u$ and $S_d$ with \qsq{} appears to be quite linear.
It is interesting also that the slope associated with the $d$ quark
is about six times larger than that of the $u$ quark.
When we consider the matrix elements of $\overline{u}\gamma_{\mu}u$ and $\overline{d}\gamma_{\mu}d$
individually, the relationship between the Pauli and the Dirac
amplitudes is quite different from when we consider
the sum of the amplitudes that results in the full hadronic matrix element (Eq.~2).   
\begin{figure}[!hbt]
\includegraphics[trim = 5mm 5mm 0mm 2mm, width=0.45\textwidth]{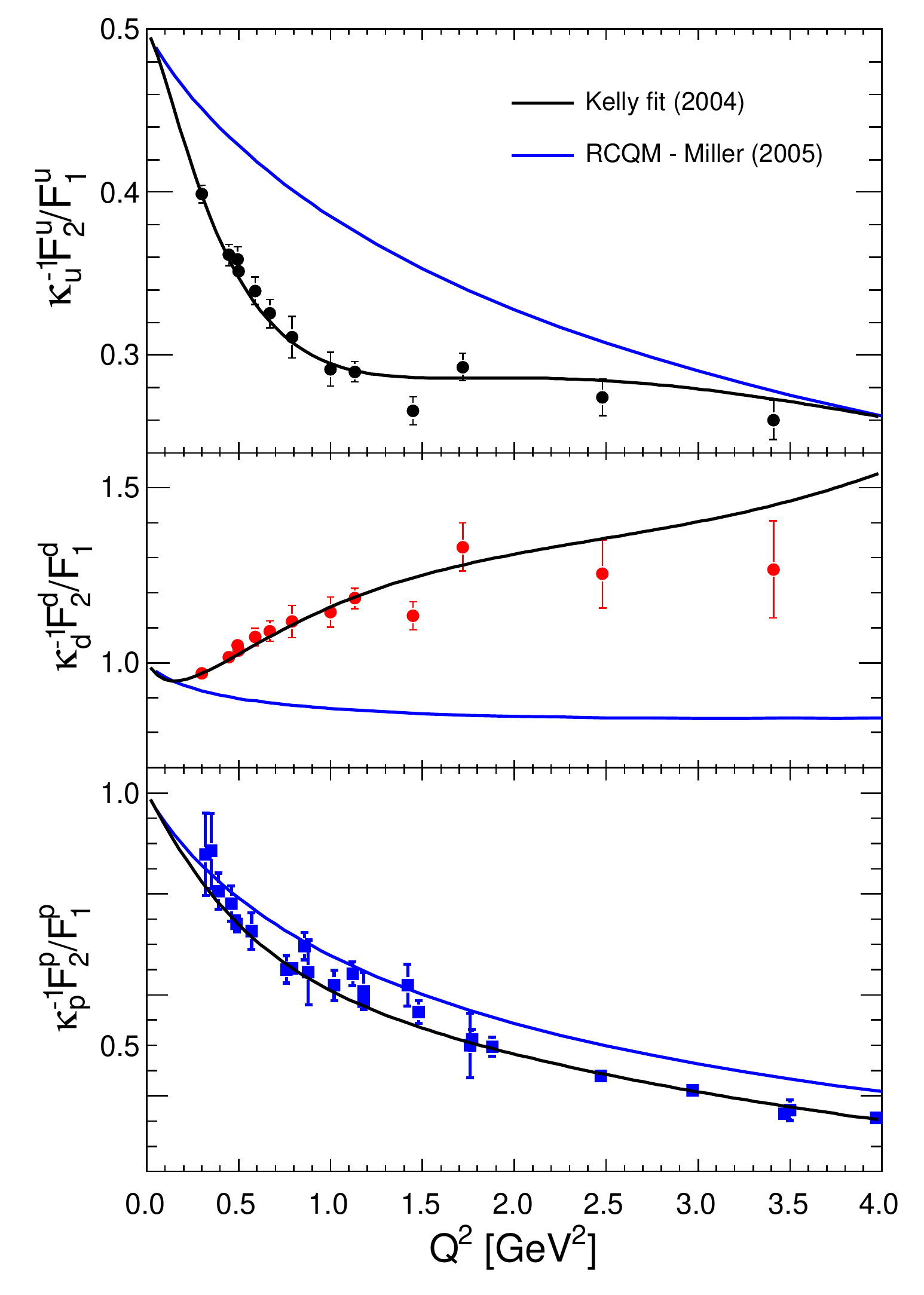}
\vskip -.05in
\caption{The ratios $\kappa_d^{-1}$\Ftwod/\Foned,  $\kappa_u^{-1}$\Ftwou/\Foneu\ and 
$\kappa_p^{-1}$\Ftwop/\Fonep\ vs. momentum transfer \qsq.
The data and curves are described in the text.}
\label{fig:f2f1} 
\vskip -.25 in
\end{figure}

While it is instructive to plot $S_u$ and $S_d$ so that we can compare
them directly with the widely discussed $S_p$ for the proton,  
the inclusion of the factor of \qsq\ masks the detailed behavior as \qsq{} approaches zero.
We thus plot in the top two panels of Fig.~\ref{fig:f2f1} the quantities $\kappa^{-1}_u
\,$\Ftwou/\Foneu{} and $\kappa^{-1}_d \,$\Ftwod/\Foned.
Here, a second aspect of the behavior of the flavor decomposed 
form factors appears that is quite intriguing.   
These ratios are relatively constant for $Q^2$ greater than $\rm \sim
1\,GeV^2$, but have a more complex behavior for lower values of \qsq.  
This might be interpreted as a transition between
a region where the virtual photon coupling to the 
three-quark component in the wave function 
dominates (higher \qsq) and a region where the inclusion of
a coupling to a five-quark component is essential (lower \qsq).
We note also that the ratio \Ftwo/\Fone\ for the proton does not show
a different behavior above and below $\rm 1\,GeV^2$ as one can see
in the bottom panel of Fig.~\ref{fig:f2f1}.
The calculation of the form factors in a relativistic constituent quark model (RCQM)~\cite{mil05}
(shown by the blue curves in Fig.~\ref{fig:f2f1}) deviates considerably
from the data which illustrates  the discriminating power of the flavor separated form factors.
The empirical Kelly fit (which predates Ref.~\cite{rio10}),
corresponds to the black curves, and is in reasonable
agreement with the data, particularly at lower \qsq.

\begin{figure}[!t]
\includegraphics[trim = 8mm 8mm 0mm 0mm, width=0.50\textwidth]{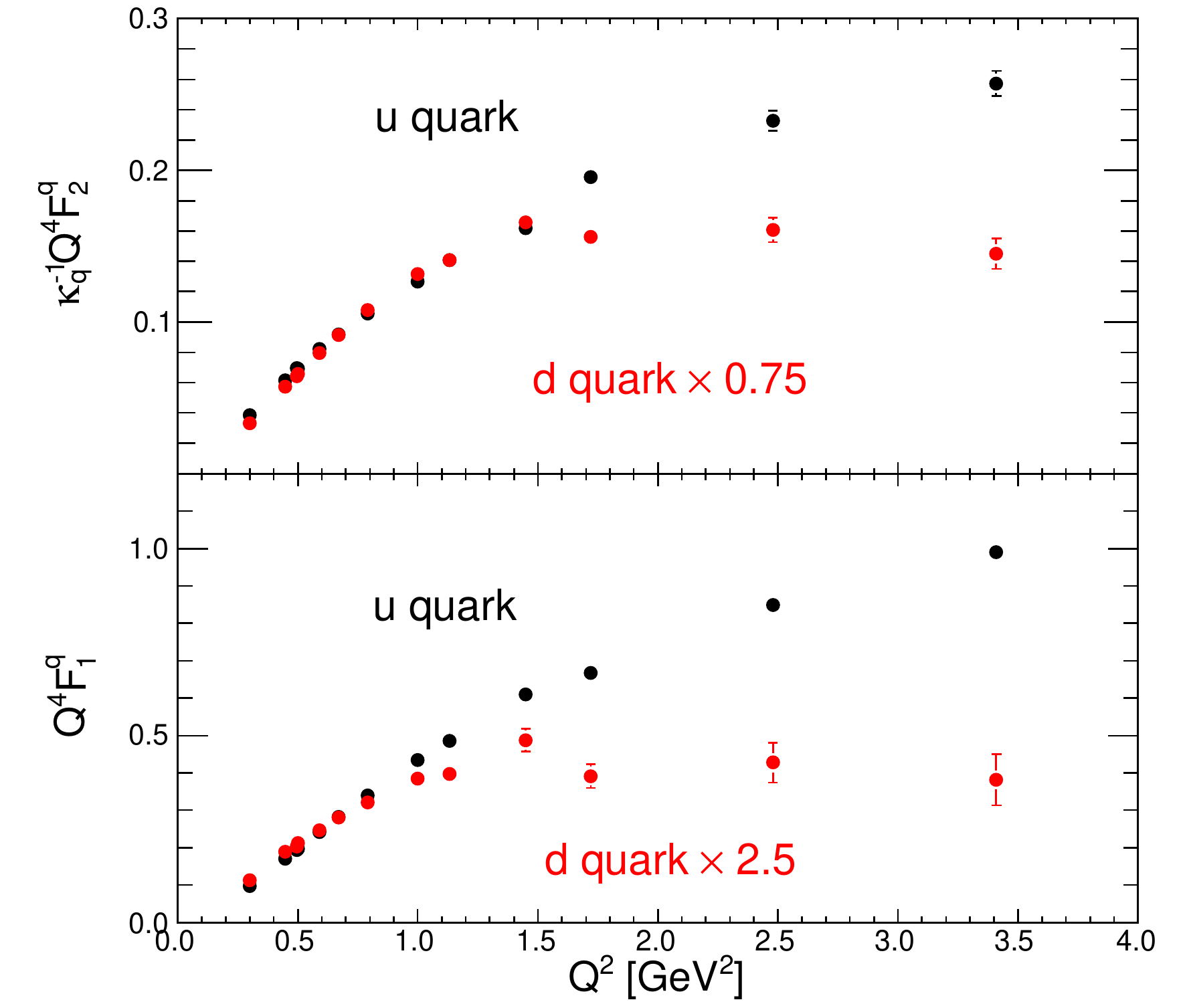}
\caption{The \qsq-dependence for the $u$- and $d$-contributions to the
proton form factors (multiplied by $Q^4$).
The data points are explained in the text.}
\label{fig:q4f1ud} 
\vskip -.2in
\end{figure}

The form factors $F_1^u$, $F_1^d$, $F_2^u$ and $F_2^u$  are shown in Fig.~\ref{fig:q4f1ud},
all multiplied by $Q^4$ for better clarity in the high-\qsq\ range.
The values are given in Table I.
\vskip -.2 in
\begin{table}[!ht]
\begin{center}
\caption{The flavor contributions to the proton form factors,
obtained using \GEn/\GMn\ form factor data from Refs.[13-18] and
the Kelly fit~\cite{kel04} for the other form factors.  
The \qsq\ values are given in \gevsq.} 
\label{tab:results}
\vspace{+0.1 cm}
\begin{tabular}{|c|c|cccc|}
\hline
~\qsq~ & Ref. & \Foneu&\Foned& \Ftwou& \Ftwod \\
\hline
0.30~&\cite{gla05}   & ~$1.075(6)$& $0.505(12)$& $0.716(6)$& $-0.995(12)$~ \\
0.45~&\cite{pla06}   & ~$0.853(6)$& $0.377(12)$& $0.515(6)$& $-0.777(12)$~ \\
0.50~&\cite{zhu01}   & ~$0.789(6)$& $0.332(12)$& $0.473(6)$& $-0.708(12)$~ \\
0.50~&\cite{war04}   & ~$0.789(4)$& $0.340(7)$ & $0.463(4)$& $-0.713(7)$~ \\
0.59~&\cite{gla05}   & ~$0.695(6)$& $0.283(13)$& $0.394(6)$& $-0.617(13)$~ \\
0.67~&\cite{ber03}   & ~$0.628(6)$& $0.249(12)$& $0.342(6)$& $-0.552(12)$~ \\
0.79~&\cite{gla05}   & ~$0.544(8)$& $0.206(15)$& $0.283(8)$& $-0.467(15)$~ \\
1.00~&\cite{war04}   & ~$0.434(5)$& $0.154(10)$& $0.211(5)$& $-0.357(10)$~ \\
1.13~&\cite{pla06}   & ~$0.379(3)$& $0.124(5)$& $0.183(3)$& $-0.298(5)$~ \\
1.45~&\cite{pla06}   & ~$0.290(3)$& $0.093(6)$& $0.128(3)$& $-0.213(6)$~ \\
1.72~&\cite{rio10}   & ~$0.2257(22)$& $0.0529(43)$& $0.1103(22)$& $-0.1429(43)$~ \\
2.48~&\cite{rio10}   & ~$0.1380(18)$& $0.0278(35)$& $0.0632(18)$& $-0.0707(35)$~ \\
3.41~&\cite{rio10}   & ~$0.0851(12)$& $0.0131(24)$& $0.0370(12)$& $-0.0337(24)$~ \\
\hline
\end{tabular}
\end{center}
\vspace{-0.20 cm}
\end{table}

Up to \qsq{} $\approx$ 1~\gevsq\ there is a constant scaling
factor of $\sim$2.5 for \Fone\ and $\sim$0.75 for \Ftwo, between the $u$- and $d$-quark contributions.
Above 1~\gevsq\ the $d$-quark contributions to both
nucleon form factors multiplied by $Q^4$ become constant in contrast to
the $u$-quark contributions which continue to rise.
These experimental results are in qualitative agreement with the predictions 
for the moments of the generalized parton distributions reported in Ref.~\cite{die05}.
It is interesting to note that the $d$-contributions correspond to the flavor that is represented singly
in the proton, whereas the $u$-contributions correspond to the flavor for which there are two quarks.
In the framework of Dyson-Schwinger equation calculations, the reduction 
of the ratios \Foned/\Foneu\  and \Ftwod/\Ftwou\ at high \qsq\ is related to diquark degrees of freedom~\cite{rob07}.
The reduction of these ratios has the immediate consequence
that $S_p$ has its observed shape despite the fact that  $S_u$ and $S_d$ are
almost linear with \qsq.

Another representation of the Dirac form factor is the infinite momentum frame density, 
$\rho_{_D}$, given by the expression $\rho_{_D}(b)=\int ( {Q dQ}/{2\pi}) J_{_0}(Qb)F_1(Q^2)$~\cite{mil07},
where $J_{_0}$ is the zeroth order Bessel function and $b$ is the impact parameter.
The faster drop off of the $d$-quark form factors in Fig.~\ref{fig:q4f1ud} implies that
the $u$ quarks have a significantly tighter distribution than the $d$
quarks in impact-parameter space, as was noticed in Ref.~\cite{mil08}.

In summary, we have performed a flavor separation of the elastic
electromagnetic form factors of the nucleon.
We find that for large \qsq\ the $d$-quark contributions to both proton
form factors are reduced relative to the $u$-quark contributions.
We find also that the \qsq-dependencies of  the flavor-decomposed 
quantities $S_u$ and $S_d$ are relatively linear in contrast to the more 
complicated behavior of $S_p$ and $S_n$.  This linearity
is due to the fact, as yet unexplained,  that the ratios $F_2^u/F_1^u$ and 
$F_2^d/F_1^d$ are constant within experimental errors for \qsq~$>1$~\gevsq.
At \qsq~$<1$~\gevsq, however, these same ratios show significant variation.  
Given the linearity of $S_u$ and $S_d$, it is quite clear that the precocious scaling
of the proton form factors and the consistency of the proton data with
the updated pQCD description of Ref.~\cite{bel03} are the result of 
the different behaviors of the  $u$- and $d$-quark contributions to the
proton form factors.
Further measurements of \GEn/\GMn~\cite{gen2} 
will allow us to extend the flavor decomposition to \qsq=10~\gevsq\
and to explore the \qsq\ range over which 
the apparent constant  behavior  of \Ftwou/\Foneu\ and \Ftwod/\Foned\ persists.

\begin{acknowledgments}
This work was supported by the U.S. Department of Energy. 
Jefferson Science Associates, LLC, operates Jefferson Lab for 
the U.S. DOE under U.S. DOE contract DE-AC05-060R23177.
\end{acknowledgments}

\end{document}